\documentclass[twocolumn,aps,showpacs,prl,amsmath,amssymb,floatfix,superscriptaddress]{revtex4}
\usepackage{graphicx}
\usepackage{dcolumn}
\usepackage{bm}
\usepackage{array}
\usepackage{float}
\usepackage{supertabular}
\usepackage{longtable}
\begin{document}

\title{Voter Model on Heterogeneous Graphs}
\author{V.~Sood}\email{vsood@bu.edu}
\author{S.~Redner}\email{redner@cnls.lanl.gov}\altaffiliation{On leave of absence from Department of 
Physics, Boston University}
\affiliation{Theory Division  and Center for Nonlinear Studies, Los Alamos National Laboratory, 
Los Alamos, New Mexico 87545}

\begin{abstract}
  We study the voter dynamics model on heterogeneous graphs.  We exploit the
  non-conservation of the magnetization to characterize how consensus is
  reached.  For a network of $N$ nodes with an arbitrary but uncorrelated
  degree distribution, the mean time to reach consensus $T_N$ scales as
  $N\mu_1^2/\mu_2$, where $\mu_k$ is the $k^{\rm th}$ moment of the degree
  distribution.  For a power-law degree distribution $n_k\sim k^{-\nu}$,
  $T_N$ thus scales as $N$ for $\nu>3$, as $N/\ln N$ for $\nu=3$, as
  $N^{(2\nu-4)/(\nu-1)}$ for $2<\nu<3$, as $(\ln N)^2$ for $\nu=2$, and as
  ${\cal O}(1)$ for $\nu<2$.  These results agree with simulation data for
  networks with both uncorrelated and correlated node degrees.

\end{abstract}


\maketitle

In this letter we study the voter model \cite{L99} on heterogeneous networks
and show that its behavior is dramatically different than that on regular
lattices.  Many recent studies of basic statistical mechanical models on
heterogeneous graphs have begun to elucidate how the dispersity in node
degree (the number of links attached to a node) affects critical behavior.  A
representative but incomplete set of examples include percolation
\cite{MN00}, the Ising model \cite{DGM02,B02,H02,H03,LPST04}, diffusion and
random walks \cite{AKS,NR,MK,SR04}, as well as the voter model itself
\cite{CVV03,VC04,SEM04,WH04}.

The voter model is perhaps the simplest and most completely solved example of
cooperative behavior.  For these reasons, our analytical predictions for the
voter model on heterogeneous networks should provide new insights into the
role of underlying heterogeneity on dynamical cooperative behavior.  In the
model, each node of a graph is endowed with two states -- spin up and spin
down.  The evolution consists of: (i) picking a random voter; (ii) the
selected voter adopts the state of a randomly-chosen neighbor.  These steps
are repeated until a finite system necessarily reaches consensus.

One basic property of the voter model is the exit probability, namely, the
probability that the system ends with all spins up, $E_+(\rho_0)$, as a
function of the initial density of up spins $\rho_0$.  Because the mean
magnetization (averaged over all realizations and all histories) is conserved
on regular lattices, and because the only possible final states are
consensus, $E_+(\rho_0)=\rho_0$ \cite{L99}.  A second basic property is the
mean time to reach consensus, $T_N$.  For regular lattices in $d$ dimensions,
it is known that $T_N$ scales with the number of nodes $N$ as $N^2$ in $d=1$,
as $N\ln N$ in $d=2$, and as $N$ in $d>2$ \cite{L99,K02}.  For heterogeneous
networks, we find that $T_N$ grows as $N\mu_1^2/\mu_2$, where $\mu_k$ is the
$k^{\rm th}$ moment of the degree distribution of the network
[Eq.~(\ref{TRHS})].  In contrast to lattice systems, the $N$ dependence of
$T_N$ is generally sublinear.

To understand how dispersity in node degree affects voter model dynamics, we
first examine the illustrative example of the complete bipartite graph.  We
then extend this approach to determine the behavior of the voter model on
networks with power-law degree distributions, but with no correlations
between node degrees.  Finally, we validate our theoretical results by
simulations of the voter model on networks with power-law degree
distributions, both with and without node degree correlations.

\begin{figure}[ht]
  \vspace*{0.cm}
  \includegraphics*[width=0.25\textwidth]{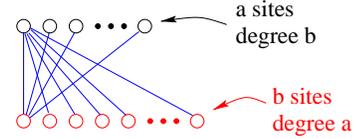}
 \caption{The complete bipartite graph $K_{a,b}$.   }
 \label{Kab}
\end{figure}

Consider the voter model on the complete bipartite graph $K_{a,b}$ of $N=a+b$
nodes that are partitioned into two subgraphs $a$ and $b$ (Fig.~\ref{Kab}).
Each node in the $a$ subgraph is connected to all nodes in the $b$ subgraph,
and vice versa.  Let $N_{a,b}$ be the respective number of up spins on each
subgraph.  In an update event, these numbers change according to
\begin{eqnarray}
\label{RE-N}
dN_a &=& \frac{a}{a+b}\Big[\frac{a-N_a}{a}\frac{N_b}{b}-\frac{N_a}{a}\frac{b-N_b}{b}\Big]\,, \nonumber\\
dN_b &=& \frac{b}{a+b}\Big[\frac{b-N_b}{b}\frac{N_a}{a}-\frac{N_b}{b}\frac{a-N_a}{a}\Big]\,.
\end{eqnarray} 
For $dN_a$, the gain term accounts for flipping a down spin in subgraph $a$
due to its interaction with an up spin in $b$, while the loss term accounts
for flipping an up spin in subgraph $a$.  The second equation accounts for
the evolution of $N_b$.  Since the time increment for an event is
proportional to $1/(a+b)$, the subgraph densities $\rho_a=N_a/a$ and
$\rho_b=N_b/b$ obey $\dot\rho_{a,b} = \rho_{b,a}-\rho_{a,b}$, with solution
\begin{eqnarray}
\label{rhoav}
\rho_{a,b}(t)=\frac{1}{2}[\rho_{a,b}(0)-\rho_{b,a}(0)]\, e^{-2t}+ \frac{1}{2}[\rho_a(0)+\rho_b(0)].
\end{eqnarray}

While the sum of the subgraph densities, $\rho_a+\rho_b$ is conserved, the
magnetization $m=(a\rho_a+b\rho_b)/(a+b)$ is not \cite{SEM04}.  However, the
bias in the rate equations for $\rho_a$ and $\rho_b$ drive the subgraph
densities to the common value $\rho_{a,b}(\infty)= \frac{1}{2}
[\rho_a(0)+\rho_b(0)]$ and magnetization conservation is restored as this
final state is approached.  It also bears mentioning that the magnetization
itself is conserved if the update rule is link-based \cite{SEM04}.

Since $\rho_a+\rho_b$ is conserved, the sum of the subgraph densities in the
final state equals 2 with probability $E_+$.  Thus the exit probability is
\begin{eqnarray}
\label{EP}
E_+  =  \frac{1}{2}[\rho_a(0) + \rho_b(0)].
\end{eqnarray}
When the initial spins on the two subgraphs are oppositely oriented, there is
an equal probability of ending with all spins up or all spins down, {\it
  independent\/} of the subgraph sizes.  In the extreme case of the star
graph $K_{a,1}$, with $a\gg 1$ up spins at the periphery and a single down
spin at the center, there is only a 50\% change that the system ends with all
spins up.

We now study the mean time until consensus $T_N(\rho_a,\rho_b)$ -- either all
spins up or all spins down -- as a function of $N$, $\rho_a$, and $\rho_b$.
This consensus time satisfies the recursion formula \cite{K97,fpp}:
\begin{eqnarray}
\label{T-REC-BIP}
T_N(\rho_a,\rho_b)&\!=\!&\nonumber
{\bf P}(\rho_a,\rho_b\to\rho_a\!\pm\!\frac{1}{a},\rho_b)
[T_N(\rho_a\!\pm\!\frac{1}{a},\rho_b)+\delta t]\nonumber\\
&\!+\!&{\bf P}(\rho_a,\rho_b\to\rho_a,\rho_b\!\pm\!\frac{1}{b})
[T_N(\rho_a,\rho_b\!\pm\!\frac{1}{b})+\delta t]\nonumber\\
&\!+\!&{\bf P}(\rho_a,\rho_b\to\rho_a,\rho_b),
[T_N(\rho_a,\rho_b)+\delta t],
\end{eqnarray}
where $\delta t=1/(a+b)\equiv 1/N$ is the time step for a single spin-flip
attempt.  For example, the first term (a shorthand for two contributions)
accounts for flipping a down (up) spin in subgraph $a$ so that
$\rho_a\to\rho_a\pm \frac{1}{a}$.  The probability for flipping a down spin
in subgraph $a$ is ${\bf P}(\rho_a,\rho_b\to\rho_a+\frac{1}{a},\rho_b) =
\frac{a}{a+b}\,(1-\rho_a)\rho_b$, where $\frac{a}{a+b}(1-\rho_a)$ is the
probability to choose a down spin in subgraph $a$ and $\rho_b$ is the
probability to choose an up spin in subgraph $b$.  This equation is subject
to the boundary conditions $T_N(0,0)=T_N(1,1)=0$.

\begin{figure}[ht]
\vspace*{0.cm}
\includegraphics*[width=0.475\textwidth]{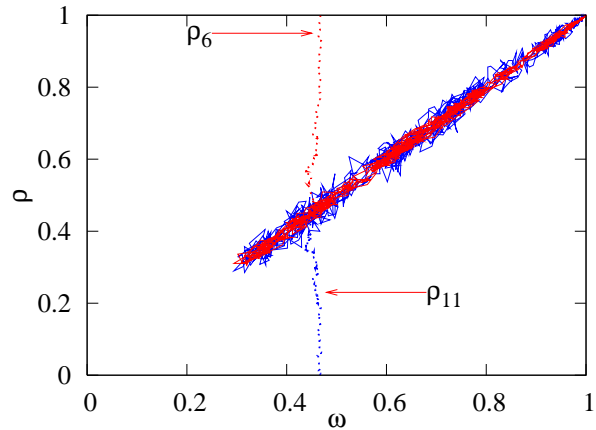}
\caption{Trajectories of  $\rho_{6}(t)$ (degree less than
  $\mu_1$) and $\rho_{11}(t)$ (degree greater than $\mu_1=8$) versus
  $\omega$, for one realization of the voter model on a network of $2 \times
  10^5$ nodes, with degree distribution $n_k\sim k^{-2.5}$.  The initial
  state is $(\rho_{k>\mu_1}(0),\rho_{k\leq\mu_1}(0)) =(0,1)$.  The dotted
  curves are the initial transient for $t\alt 1$, after which diffusive
  motion leads to consensus at $(1,1)$.}
\label{BGEvo}
\end{figure}

Expanding this recursion formula to second order, we find, after
straightforward algebra,
\begin{eqnarray}
\label{Bip-CT}
N\delta t &=& (\rho_a-\rho_b)(\partial_a-\partial_b)T_N(\rho_a,\rho_b) \\
&& - \frac{1}{2}(\rho_a+\rho_b-2\rho_a\rho_b)\left( \frac{1}{a}\partial_a^2 +
 \frac{1}{b}\partial_b^2\right)T_N(\rho_a,\rho_b)\nonumber
\end{eqnarray}
where $\partial_i$ denotes a partial derivative with respect to $\rho_i$.
The first term on the right accounts for a convection that drives the system
to equal subgraph magnetizations in a time of order one.  Subsequently,
diffusive fluctuations govern the ultimate approach to consensus (see
Fig.~\ref{BGEvo}).
We thus compute the consensus time by replacing the subgraph densities
$\rho_a$ and $\rho_b$ by their common value $\rho$.  In so doing, we ignore
the initial transient for $t\sim O(1)$, during which the subgraph densities
are unequal.  We also transform the derivatives with respect to $\rho_a$ and
$\rho_b$ in Eq.~(\ref{Bip-CT}) to derivative with respect to $\rho$ to yield
\begin{equation}
\label{t-eqn}
\frac{1}{4}\rho(1-\rho)\left(\frac{1}{a}+\frac{1}{b}\right)\partial^2 T_N = -1,
\end{equation}
with solution
\begin{equation}
T_{N}(\rho) = -\frac{4 a b}{a+b}\left[ (1\!-\!\rho)\ln(1\!-\!\rho) + \rho\ln\rho\right]
\end{equation}
Notice that if $a = {\cal O}(1)$ and $b = {\cal O}(N)$ (star graph), the
consensus time $T_{N}\sim{\cal O}(1)$, while if both $a$ and $b$ are ${\cal
  O}(N)$, then $T_{N}\sim {\cal O}(N)$, as on a complete graph.

We now extend this approach to graphs with arbitrary degree distributions but
without degree correlations, {\it i.e.}, we treat all nodes with the same
degree as equivalent \cite{PS}.  We define $\rho_k$ as the density of up
spins in the subset of nodes of degree $k$.  Similar to
Eq.~(\ref{T-REC-BIP}), the recursion for the mean consensus time on a
heterogeneous graph, with initial densities $\{\rho_k\}$, is:
\begin{eqnarray}
\label{T-REC-HET}
T_N(\{\rho_k\}) &=& \sum_k {\bf P}(k; \mp \to \pm)
[T_N(\rho_k\pm \delta_k)+\delta t]\nonumber\\
&+& \sum_k {\bf Q}(\{\rho_k\}) [T_N(\{\rho_k\})+\delta t],
\end{eqnarray}
where ${\bf P}(k;\mp \to \pm)$ is the probability that a spin down (up) on a
node of degree $k$ flips in an update, ${\bf Q}$ is the probability of no
flip, and $\delta_k=1/(N n_k)$ is the change in $\rho_k$ when a spin flip
occurs at a site of degree $k$.  Here $n_k$ is the fraction of sites with
degree $k$.

Since the probability of choosing a node is $1/N$, the spin flip probability
may be written as
\begin{eqnarray}
\label{FLIP-PROB-DU}
{\bf P}(k; \mp \to \pm) &=&  \sum_{x:\left({k_x = k}\atop{ s_x=\mp}\right)}\frac{1}{N} 
\,\,\sum_{{y:s_y=\pm}}\,\frac{1}{k}{\rm A}_{x y}\,,
\end{eqnarray}
where $A_{xy}$ is the adjacency matrix element between nodes $x$ and $y$
($A_{xy}=1$ if $x$ and $y$ are connected and $A_{xy}=0$ otherwise).  Thus the
second sum is the probability that a node $x$ with degree $k$ chooses a
neighbor with spin up (down).  Under the mean-field assumption that
neighboring node degrees are uncorrelated, we write $A_{xy}$ as ${k_x
  k_y}/{\mu_1 N}$, where $\mu_1\equiv \sum_k k n_k$ is the mean node degree
for the graph.  That is, we replace ${\rm A}_{x y}$ by the probability that
an edge between node $x$ of degree $k_x$ and node $y$ of degree $k_y$ exists.
Then the second sum in Eq.~(\ref{FLIP-PROB-DU}) for spin up and spin down
simplifies respectively to
\begin{eqnarray*}
\frac{1}{\mu_1 N}\sum_{{y}\atop{s_y=+}}  k_y &=& \frac{1}{\mu_1} \sum_{j} j n_{j}\rho_{j}\equiv \omega\\
\frac{1}{\mu_1 N}\sum_{{y}\atop{s_y=-}} k_y &=& \frac{1}{\mu_1} \sum_{j} j
 n_{j}(1-\rho_{j})\equiv 1-\omega.
\end{eqnarray*}
Namely, we decompose the nodes $y$ according to their degree, and we define
$\omega$ as the average degree-weighted density of up spins.  In this
formulation, each spin of given sign flips with the same probability that is
a function of the degree-weighted magnetization rather than of the global
magnetization, as in the case for degree-regular graphs.  Since the first sum
in Eq.~({\ref{FLIP-PROB-DU}}) gives the density of down (up) spin in the
subset of nodes with degree $k$, we now write ${\bf P}(k; - \to +) =
n_k\omega (1 - \rho_k)$, and similarly, ${\bf P}(k; + \to -) = n_k(1-\omega)
\rho_k$.  Finally, the probability that there is no change in a single spin
flip attempt is ${\bf Q}(\{\rho\}) = 1-\sum_k \left({\bf P}(k;-\to+)+{\bf
    P}(k;+\to-)\right)$.

These simplifications enable us to write Eq.~(\ref{T-REC-HET}) as
\begin{eqnarray}
\label{TRH}
-\delta t 
&\!=\!&\sum_k n_k\left[\omega(1\!-\!\rho_k) (T(\rho_k\!+\!\delta_k)-T(\{\rho_k\}))\right]\nonumber\\
&\!+\!&\sum_k n_k\left[(1\!-\!\omega)\rho_k (T(\rho_k\!-\!\delta_k)-T(\{\rho_k\}))\right]
\end{eqnarray}
Expanding this recursion to second order we obtain
\begin{eqnarray}
\label{TRHDEQ}
 N\delta t=\sum_k (\rho_k\!-\!\omega)\partial_{k} T -
\sum_k \frac{(\omega\!+\!\rho_k\!-\!2\omega\rho_k)}{2Nn_k}\,\,
\partial^{2}_{k} T,
\end{eqnarray}
where $\partial_k$ denotes the partial derivative with respect to $\rho_k$.
The convective terms on the right-hand side again drive the system to the
state where $\rho_k$ is equal to the weighted magnetization $\omega$ for all
$k$.  

To check this convective behavior, we followed the evolution of single
realizations of the voter model on scale-free graphs with degree distribution
$n_k \propto k^{-2.5}$ and mean degree $\mu_1 = 8$ generated according to the
Molloy-Reed (MR) model \cite{MR}.  Each node is assigned a random number of
stubs $k$ that is drawn from a specified degree distribution.  Pairs of
unlinked stubs are then randomly joined.  This construction eliminates degree
correlations between neighboring nodes.  For the initial state, we assign all
nodes with degree larger than $\mu_1$ as spin down and all remaining nodes as
spin up.  A plot of the spin up densities $\rho_{6}$ and $\rho_{11}$ for
nodes of degrees 6 and 11 versus the degree-weighted up-spin density shows
that these ``subgraph'' densities quickly approach equal values
(Fig.~\ref{BGEvo}).  Analogous behavior occurs on the bipartite graph and on
scale-free networks with degree correlations.

For long times, we thus drop the convective terms and set $\rho_k = \omega~~
\forall k$.  Concomitantly, we transform the partial derivatives with respect
to $k$ to derivatives with respect to $\omega$ by using $\partial_k\omega =
n_k k/\mu_1$ to reduce (\ref{TRHDEQ}) to
\begin{equation}
\frac{1}{N}\sum_k\left(\frac{k^2}{\mu_1^2}n_k\right)\omega(1-\omega)\,\,\partial^2_\omega T =-1.
\end{equation}
Since $\sum_k k^2 n_k=\mu_2$, the second moment of the degree distribution,
this equation can be reduced to a similar form to (\ref{t-eqn}), with
solution
\begin{equation}
\label{TRHS}
T_N(\omega) = -N \frac{\mu_1^2}{\mu_2}\,\,[(1-\omega)\ln(1-\omega)+\omega\ln\omega\,].
\end{equation}

For a scale-free network \cite{BA} with degree distribution $n_k\sim
k^{-\nu}$, the $m^{\rm th}$ moment is $\mu_m\sim \int ^{k_{\rm max}} k^m
n_k\, dk$.  Here $k_{\rm max}\sim N^{1/(\nu-1)}$ is the maximal degree in a
finite network of $N$ nodes; this is obtained from the extremal condition
$\int_{k_{\rm max}} k^{-\nu}\, dk = N^{-1}$ \cite{KR02}.  Thus the second
moment diverges at the upper limit for $\nu\leq 3$ while the first moment
diverges for $\nu\leq 2$.

Assembling the results for the moments, the mean consensus time on a
scale-free graph has the $N$ dependence
\begin{equation}
\label{Tf}
T_N\sim
\begin{cases}
N  & \nu>3,\cr
N/\ln N & \nu=3,\cr
N^{(2\nu-4)/(\nu-1)} & 2<\nu<3,\cr
(\ln N)^2 & \nu=2,\cr
{\cal O}(1) & \nu<2.
\end{cases}
\end{equation}
The prediction $T_N\sim N/\ln N$ for $\nu=3$ may explain the apparent
power law $T_N\sim N^{0.88}$ reported in a previous simulation of the voter
model on such a network \cite{SEM04}.

\begin{figure}[ht]
  \vspace*{0.cm}\hspace*{-0.3in}
 \includegraphics*[width=0.52\textwidth]{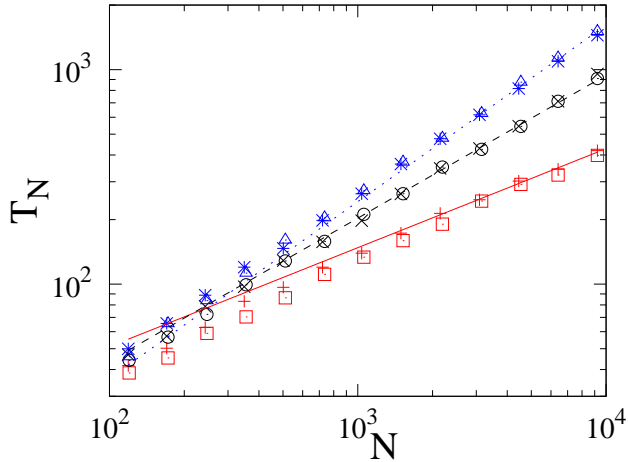}
\caption{Consensus time $T_N$ versus $N$ on scale-free Molloy-Reed networks 
  with degree distribution $n_k = k^{-\nu}$ for $\nu = 2.3\ (+)$,
  $2.5(\times)$, and $2.7\ (\ast)$.  Also shown are corresponding results for
  the GNR (open symbols).  Only every second data point is shown.  The data
  are all based on 10 graph realizations, each with 100 voter model
  realizations.  The lines are the expected asymptotic slopes.}
 \label{TCU}
\end{figure}

To test our predictions, we simulated the voter model on the Molloy-Reed (MR)
network \cite{MR} and on the growing network with redirection (GNR)
\cite{KR01}.  The GNR is built by adding nodes sequentially, where each new
node attaches either to a randomly-selected node with probability $1-r$ or to
the ancestor of this target with probability $r$.  We chose the out degree of
each node to be 4, and redirection was applied to each outgoing link of the
new node.  This construction gives a network with a power-law degree
distribution $n_k\propto k^{-\nu}$, with $\nu=1+\frac{1}{r}$ in the range
$(2,\infty)$ as $r$ is varied between 0 and 1.

Fig.~\ref{TCU} shows the $N$ dependence of $T_N$ for representative values of
the degree exponent $\nu$ for both the MR network and the GNR.  The results
for the two networks with the same $\nu$ are extremely close, suggesting that
degree correlations have a small effect on voter model dynamics.  There is
also curvature in the date that originates from finite-$N$ effects.  Using
the maximal degree $k_{\rm max}\sim N^{1/(\nu-1)}$ in the definition of the
moments ultimately lead to the exponent for $T_N$ being modified by the
corrections, for $\nu$ between 2 and 3,
\begin{eqnarray}
\label{Ecorr}
\frac{d{\rm ln}T}{d{\rm ln}N} = \frac{2\nu\!-\!4}{\nu\!-\!1}\left(1 
-a\,N^{\frac{2-\nu}{\nu-1}} + b\, N^{\frac{\nu-3}{\nu-1}}\right),
\end{eqnarray}
where $a$ and $b$ are of order 1.

For $\nu$ close to 2 or 3, the leading correction term decays slowly in $N$,
causing a discrepancy between our numerics and the theory.  For example, for
$\nu=2.3$ in Fig.~\ref{TCU}, the numerical best-fit slope to the data
decreases from 0.53 to 0.48 as we successively eliminate the first 18 data
points.  This accords well with the theoretical prediction of 0.46 for the
slope from Eq.~(\ref{Tf}).  For $\nu=2.5$, the two correction terms both
decay at the same rate and have opposite sign.  Here we may expect the
smallest corrections, as borne out by the data -- the best-fit slope
decreases from 0.680 to 0.671 as the first 18 data points are deleted, while
the theoretical prediction for the slope is 2/3.  The case $\nu=2.7$ has the
slowest-decaying correction term and here we observe the largest deviation
between simulation and theory -- the slope remains in the range 0.77 -- 0.79
as the first 18 points are deleted, while theory predicts a slope of 0.82.

In summary, the voter model on heterogeneous networks approaches consensus by
a two-stage process of quick evolution to a opinion-homogeneous state
followed by a diffusive evolution to final consensus.  By neglecting node
degree correlations, the consensus time $T_N$ on scale-free graphs has the
following dependence on the degree distribution exponent $\nu$: for $\nu<2$,
$T_N\sim {\cal O}(1)$, while for $\nu>3$, $T_N\sim N$.  In the intermediate
regime of $2<\nu<3$, $T_N\sim N^{(2\nu-4)/(\nu-1)}$.  Generically, $T_N$
grows sublinearly with $N$; that is, high-degree nodes greatly accelerate the
approach consensus.  Finally, the $N$-dependence of $T_N$ is virtually the
same for networks without and with degree correlations.

\acknowledgments{We gratefully acknowledge financial support from NSF grants
  DMR0227670 (at BU) and DOE grant W-7405-ENG-36 (at LANL).  }

\end{document}